\documentclass{jca}
\usepackage{tikz}
\usetikzlibrary{arrows}

\usepackage[utf8x]{inputenc}
\usepackage{amsmath,graphicx}

\usepackage[draft,firstpage]{draftmark}
\draftmarksetup{fontfamily=ptm,angle=0,scale=0.07,mark={\mbox{Published in: \emph{Journal of Cellular Automata}, 10(1--2):149--160, 2015.}},color=red,xcoord=-78,ycoord=120}

\usepackage{url}
\usepackage{multirow}

\newtheorem{proposition}{Proposition}

\begin{document}
\title{Solving two-dimensional density classification problem with two probabilistic cellular automata}
\author{Henryk Fuk\'s}
\institute{Brock University\\
Department of Mathematics and Statistics\\
St. Catharines, ON, Canada\\
Email: hfuks@brocku.ca}

% 
% \IEEEauthorblockN{Henryk Fuk\'s}
% \IEEEauthorblockA{
% Brock University\\
% Department of Mathematics and Statistics\\
% St. Catharines, ON, Canada\\
% Email: hfuks@brocku.ca}
% }

\maketitle              % typeset the title of the contribution

\begin{abstract}
The density classification problem is one of the simplest yet non-trivial computing
tasks which seem to be ideally suitable for cellular automata (CA). Unfortunately,
there exists no one-dimensional two-state CA which classifies binary
strings according to their densities. If, however, in place of simple cells
one uses agents which change their behaviour from one rule to another after a fixed number of iterations,
the classification can be performed by  the traffic rule 184 
and the majority rule 232.
This two-rule solution cannot be easily generalized to two (or higher) dimensions, because 
it critically depends on a kinetic phase transition occurring in the rule 184.
No  rule exhibiting analogous transition is known in two dimensions, most likely because no such rule 
exists. We propose, therefore,  to approach this problem form a slightly different angle,
namely by introducing a stochastic component into each of the two rules.
If one precedes each iteration of rule 184 by the stochastic ``lane changing
rule'', and each iteration of rule 232 by the stochastic ``crowd avoidance'' rule,
in the limit of infinitely many iterations the classification can be performed correctly with probability 1.
This solution can be described either in the language of CA, or using the paradigm
of agents which move and proliferate on the 2D lattice, following probabilistic rules.
\end{abstract}

\section{Introduction}
%Here: \cite{Wolf2008,Fates2d-2011,JimenezMorales2001571,EJP2325,oliv2013}.
The density classification problem (DCP) asks for a construction of a cellular automaton
rule which, when applied to a binary string of density $\rho$, converges to all
ones when $\rho>1/2$ and all zeros when $\rho<1/2$.
From the time when Gacs, Kurdyumov and Levin proposed this problem and its first
approximate solution \cite{Gacs78}, a lot of research effort went into studying
of this topic.  After it has  been  proved by Land and Belew  \cite{land95} that
the perfect two-state rule performing this task does not exist, approximate
solutions have been constructed using a variety of methods. The best currently known
solutions, found by evolutionary algorithms, are described in \cite{Wolf2008}.
For a good review of the density classification problem and its various modifications 
the reader may consult \cite{oliv2013} and references therein.

Here we shall concentrate on a two-dimensional version of DCP. For an infinite
two-dimensional array, the so-called Toom's rule is know to be a solution \cite{EJP2325},
 but its performance on finite lattices is unsatisfactory \cite{Fates2d-2011}.
Approximate solutions do exist, and similarly as in the one-dimensional case,
they have been obtained by evolutionary algorithms \cite{JimenezMorales2001571,Wolf2008}.

Since no single rule solving DCP in one dimension exists, the author
proposed  a generalized version of the problem involving two rules
and found its exact solution~\cite{Fuks97}. Later, other two-rule solutions
have been found as well~\cite{oliv2005}.  The purpose of this paper is to explore
a generalization of the two-rule solution to two dimensions. As we will shortly see,  although a 
direct and naive generalization does not work, it can be fixed
 by adding stochastic components to both rules. 
\section{Solution in one dimension}
In one dimension, in order to classify a binary string of length $L$, one needs to apply
rule 184 for a number of time steps, and then switch to rule 232 . To be more precise,
let us define $\mathbf{s}=\{s_0,s_1,\ldots, s_{L-1}\}$ to be a string of binary numbers. 
We impose periodic boundary conditions on it, so that all indices are taken modulo $L$. 
Density of $\mathbf{s}$ is defined as $\rho=\sum_{i=0}^{L-1} s_i /L$. 
We further define $R_{184}: \{0,1\}^L \to \{0,1\}^L$ as
\begin{equation}
[R_{184}(\mathbf{s})]_{i}=s_{i-1} + s_{i} s_{i+1}
                                   - s_{i-1} s_{i}, 
 \end{equation}
and $R_{232}: \{0,1\}^L \to \{0,1\}^L$ as
\begin{equation}
[R_{232}(\mathbf{s})]_{i}=\mathrm{majority\,}
\{ s_{i-1},s_{i}, s_{i+1} \}.
 \end{equation}
In \cite{Fuks97}, the following proposition is proved.
\begin{proposition}
Let $\mathbf{s}$ be a binary string of length $L$ and density $\rho$,
 and let $n=\lfloor (L-2)/2 \rfloor$, $m=\lfloor (L-1)/2 \rfloor$. Then $R_{232}^mR_{184}^{n}(s)$
consists of only 0's  if $\rho<1/2$ and of only 1's if $\rho>1/2$. If
$\rho=1/2$, $R_{232}^mR_{184}^{n}(s)$ is an alternating sequence of
0's and 1's, i.e., $\ldots01010101\ldots$.
\end{proposition}

What makes this scheme work is a specific property of rule 184. First of all, this rule is known to be number-conserving, that is, it does not change the number of zeros or ones in the string to which it is applied.
Moreover,  if the initial string has more zeros than ones, after sufficiently many iterations of rule 184
all pairs $00$ disappear. Conversely, if there is more zeros than ones in the initial string, all pairs
$11$ eventually disappear. Rule 232, on the other hand, has the property of growing continuous clusters
of zeros in the absence of 11 pairs, and growing continuous clusters of ones in the absence of 00 pairs.

In the light of the above, it is not hard to see that the combination of rules 184 and 232 performs
perfect density classification, as described in Proposition 1. Could the same scheme be applied to two-dimensional 
binary arrays?

Before we answer this question, let us remark that rule 184, being number-conserving, can be described
using the paradigm of particles or agents, if one assumes that a site in state 1 corresponds to a cell
occupied by an agent, and site in state 0 denotes empty cell. One can then show \cite{paper10} that rule 184 is 
equivalent to the following behaviour of agents: if an agent has an empty cell on the right hand side, it moves there, otherwise it does not move. All agents move at the same time. Since this is the simplest model
of road traffic with agents being cars, rule 184 is also known as the ``traffic rule''. It is obvious that
the number of agents in this rule will always be preserved.

The second rule, rule 232, can also be described in terms of agents, as follows. If an agent
has two empty neighbouring cells, it disappears, otherwise it remains. Moreover, if an empty cell is surrounded by two agents, a new agent is born there. Birth and deaths of agents happen simultaneously on the entire lattice.

The density classification problem can be now rephrased in the language of agents. We start
with a lattice where a certain unknown number of cells is occupied by agents. We want to
equip the agents with a local rule governing their behaviour such that if more than 50\% of initial cells
are occupied, in the end all cells are occupied. If less than 50\% of cells are occupied by agents,
then in the end all cells become empty. Proposition 1 tells us that if we let the agents move following
the traffic rule for  $\lfloor (L-2)/2 \rfloor$  steps and then we let them proliferate or die
following rule 232 for  $\lfloor (L-1)/2 \rfloor$ time steps, in the end we will obtain the desired
configuration.
 
\section{Rules 184 and 232 in two dimensions}
Let us now consider binary arrays $L \times L$, with entries $s_{i,j}$, where $i,j \in \{0,1, \ldots L-1\}$.
Set of all such arrays will be denoted by $\mathcal S$.
As before, we will impose periodic boundary conditions, taking all indices $i,j$ modulo $L$.

A simple and naive generalization of Proposition 1 could involve rules 184 and 232 ``lifted'' to two
dimensions, by defining 
$R_{184}: \mathcal S \to \mathcal S$ and $R_{232}: \mathcal S \to \mathcal S$ as
\begin{align}
[R_{184}(\mathbf{s})]_{i,j}&=s_{i-1,j} + s_{i,j} s_{i+1,j}
                                   - s_{i-1,j} s_{i,j},\label{rule184formula}\\ 
[R_{232}(\mathbf{s})]_{i,j}&=\mathrm{majority\,}
\{ s_{i-1,j},s_{i,j}, s_{i+1,j} \}.
 \end{align}
Unfortunately, one cannot classify densities by applying $R_{184}$ for a number of steps and then 
switching to rule 232, as it was done in one dimension. If we apply $R_{184}$ to a two-dimensional array, 
each row will remain independent of all other rows. In particular, it may happen that in some rows
all 00 pairs are eliminated, and in other rows all 11 pairs are eliminated. As a result, both
00 and 11 will still be present when one switches to rule 232, leading to (possibly) incorrect classification.
One needs to introduce some sort of interaction between rows which would allow for transfer of 00 and 11 pairs
between rows.

The author experimented with various possibilities of inter-row interactions, but all of them turned out to be 
unsatisfactory. It seems that there exists  no deterministic CA rule in two dimensions which would
have properties analogous to rule 184 in 1D. However, having in mind recent progress on one-dimensional DCP
using probabilistic rules ~\cite{paper20,Fates2011},  injection of  some stochasticity
into dynamics presents itself as a promising possibility.
Indeed, as it turns out, if one allows probabilistic CA, two-dimensional analog
of rule 184 can be  constructed rather easily.
\section{Randomization of rule 184}
\begin{figure}
\begin{center}
 \begin{tikzpicture}[->,>=stealth',shorten >=1pt,auto,node distance=1.2cm,
 % thick,main node/.style={circle,fill=blue!20,draw,font=\sffamily\Large\bfseries}]
   thick,main node/.style={font=\bfseries}]

  \node[main node] (1)              {$\star$};
  \node[main node] (2) [right of=1] {0};
  \node[main node] (3) [right of=2] {$\star$};
  
  \node[main node] (4) [below of=1] {$\star$};
  \node[main node] (5) [right of=4] {1};
  \node[main node] (6) [right of=5] {1};
  
  \node[main node] (7) [below of=4] {$\star$};
  \node[main node] (8) [right of=7] {$\star$};
  \node[main node] (9) [right of=8] {$\star$};

  \path[every node/.style={font=\sffamily}]
           (5)  edge [bend right] node[right] {$p=0.5$} (2);
\end{tikzpicture}
\end{center}
\caption{Particle (agent) representation of rule $F_X$.} \label{Fxfig}
\end{figure}
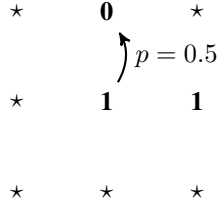

Let us first describe a rule which will be called ``lane changing rule'', illustrated in Figure~\ref{Fxfig}. 
We will call this rule $F_X$ (the meaning of the subscript $X$ will be explained later).
According to the rule $F_X$, agents move simultaneously in such a way that every agent which has another agent on the right hand side and empty site above, moves up with probability 1/2 and stays in the same place with probability 1/2. All other agents do not move. If agents were cars moving to the right, one could say that
cars which are blocked (have another car directly ahead) change lane if possible (move to the next row) -- hence
the name ``lane changing rule''.  
%%%%%%%%%%%%%%%%%%%%FX

Rule $F_X$ can of course be described formally using the language of cellular automata.
Let $X$ denote an array of independent and identically distributed
random variables $X_{i,j}$, where $i,j \in \{0,1, \ldots L-1\}$, such that
\begin{equation} \label{Xprobdist}
 \mathrm{Pr}(X_{i,j}=0)=\mathrm{Pr}(X_{i,j}=0)=\frac{1}{2}.
\end{equation}
Furthermore, let us define
\begin{equation}
 \mathbf{u}_{i,j}=\left(
\begin{array}{ccc}
s_{i-1,j+1} & s_{i,j+1} & s_{i+1,j+1}\\
s_{i-1,j} & s_{i,j} & s_{i+1,j}\\
s_{i-1,j-1} & s_{i,j-1} & s_{i+1,j-1}
\end{array}\right).
\end{equation}
Using the above notation, we define $F_X: \mathcal S \to \mathcal S$ as
\begin{equation}
[F_X(\mathbf{s})]_{i,j}=\left\{
\begin{array}{ll}
 1 & \mathrm{if\,\,}  \mathbf{u}_{i,j} =\left(
  {\setlength{\arraycolsep}{3pt} \begin{array}{lll}
  \star & \star & \star\\
  \star & 0 & \star\\
  \star & 1 & 1
  \end{array}
 }
\right) \mathrm{and\,\,} X_{i,j-1}=1,\\
 0 & \mathrm{if\,\,}  \mathbf{u}_{i,j} =\left(
  { \setlength{\arraycolsep}{3pt}
  \begin{array}{lll}
  \star & 0 & \star\\
  \star & 1 & 1\\
  \star & \star & \star
  \end{array}
  }
\right) \mathrm{and\,\,} X_{i,j}=1,\\	 
 s_{i,j}  & \mathrm{otherwise.}
\end{array}
\right.
\end{equation}

This can be written in a more compact form as
\begin{multline}\label{ruleFformula}
[F_X(\mathbf{s})]_{i,j}=
s_{{i,j}}+ \left( 1-s_{{i,j}} \right) s_{{i,j-1}}s_{{i+1,j-1}}X_{{i,j-
1}}\\-s_{{i,j}} \left( 1-s_{{i,j+1}} \right) s_{{i+1,j}}X_{{i,j}}.
\end{multline}
Using the above, one can compute the sum of array's entries after the application 
of $F_X$,  
\begin{multline} \label{ruleFsum}
\sum_{(i,j)}[F_X(\mathbf{s})]_{i,j}=\sum_{(i,j)} s_{{i,j}} 
+\sum_{(i,j)} \left( 1-s_{{i,j}} \right) s_{{i,j-1}}s_{{i+1,j-1}}X_{{i,j-
1}}\\
-\sum_{(i,j)} s_{{i,j}} \left( 1-s_{{i,j+1}} \right) s_{{i+1,j}}X_{{i,j}}.
\end{multline}
If one replaces $j$ by $j+1$ in the second sum on the right hand side, it
becomes the same as the third sum, meaning that the second and the third sum
cancel each other. This  leaves 
\begin{equation}
 \sum_{(i,j)}[F_X(\mathbf{s})]_{i,j}=
\sum_{(i,j)} s_{{i,j}}, 
\end{equation}
proving that $F$ is number-conserving.

\begin{figure}
\begin{center}
\includegraphics[scale=0.4]{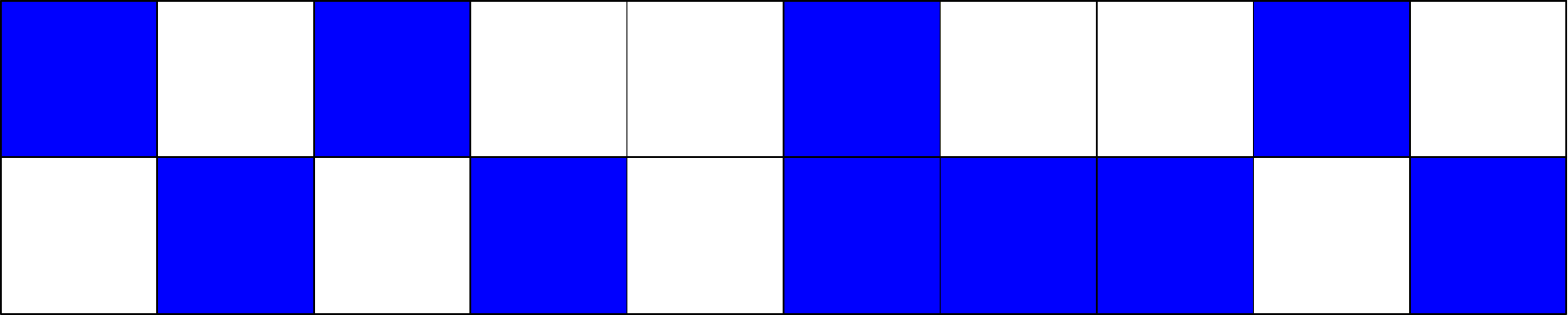}\\
(a)\\
\includegraphics[scale=0.4]{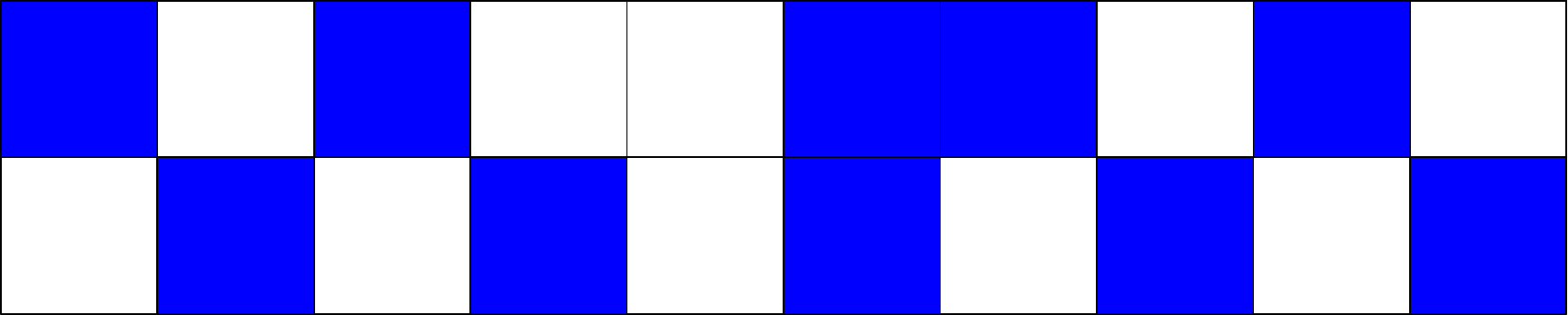}\\
(b)\\
\includegraphics[scale=0.4]{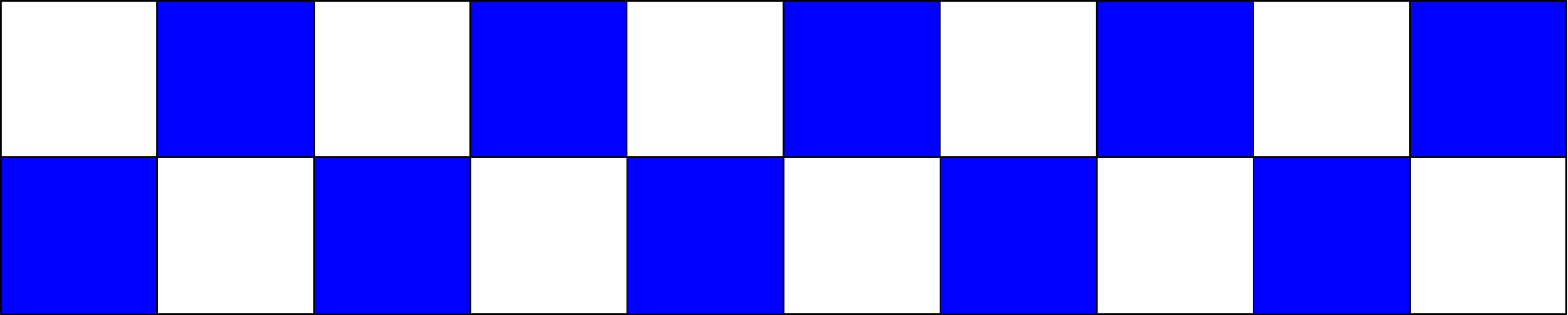}\\ 
(c)
\end{center}
 \caption{Illustration of properties of rule $F_X$: (a) initial configuration $\mathbf{s}$, 
(b) $F_X(\mathbf{s})$, (c) $R_{184} F_X(\mathbf{s})$ }
\label{fig184fx}
\end{figure}

The effect of application of $F_X$ is a ``diffusion'' of pairs 11 between rows. Figure \ref{fig184fx} illustrates
this using an initial array  with two rows only (Figure \ref{fig184fx}a). In the top row, there are two 00 pairs, and in the bottom
row, two 11 pairs. If one applied $R_{184}$ to this configuration, pairs 00 and 11 would change position
in their respective rows, but they would remain in the same row. Multiple iterations of
$R_{184}$ would have a similar effect -- pairs 00 and 11 would never be eliminated. 
Now, suppose that we apply $F_X$ first. A possible outcome of this is a configuration shown
in Figure \ref{fig184fx}b. One of the agents from the lower row (the one located at 7th position from the right)
jumped up to the upper row. Now, in the upper row we have one 00 pair and one 11 pair, and in the bottom
row neither one of them. If one now applies $R_{184}$, 00 and 11 in the top row will ``annihilate'' each other, and the resulting configuration will be as shown in Figure \ref{fig184fx}c. 

This means that instead of $R_{184}$, we need to use $F_X R_{184}$.
Rule  $F_X R_{184}$ is a probabilistic CA rule, and by combining eqs.~(\ref{ruleFformula}) and (\ref{rule184formula}),
one can obtain its explicit definition,
\begin{multline}
 F_X R_{184}(\mathbf{s})=
s_{{i-1,j}}+s_{{i,j}}s_{{i+1,j}}-s_{{i-1,j}}s_{{i,j}} \\+ 
\left( 1-s_{{i-
1,j}}-s_{{i,j}}s_{{i+1,j}}+s_{{i-1,j}}s_{{i,j}} \right)  \\ 
\times \left( s_{{i-
1,j-1}}+s_{{i,j-1}}s_{{i+1,j-1}}-s_{{i-1,j-1}}s_{{i,j-1}} \right) \\
\times \left( s_{{i,j-1}}+s_{{i+1,j-1}}s_{{i+2,j-1}}-s_{{i,j-1}}s_{{i+1,j-1}
} \right) X_{{i,j-1}} \\
- \left( s_{{i-1,j}}+s_{{i,j}}s_{{i+1,j}}-s_{{i-1
,j}}s_{{i,j}} \right)  \\
\times \left( 1-s_{{i-1,j+1}}-s_{{i,j+1}}s_{{i+1,j+1}}
+s_{{i-1,j+1}}s_{{i,j+1}} \right) \\
\times \left( s_{{i,j}}+s_{{i+1,j}}s_{{i+2
,j}}-s_{{i,j}}s_{{i+1,j}} \right) X_{{i,j}}.
\end{multline}
Note that $[F_X R_{184}(\mathbf{s})]_{i,j}$ depends not only on nearest neighbours of $\mathbf{s}_{i,j}$,
but also on second nearest neighbours. $F_X R_{184}$ is thus a probabilistic CA rule with Moore neighbourhood of range 2.

After sufficiently many iterations  of this rule,  the  binary array
will have no 00 pairs (if it had more zeros than ones at the beginning), or no 11 pairs (if it had more ones than
zeros at the beginning). This is precisely what is needed for rule 232 to do its job.
\section{Randomization of rule 232}
Once the pairs 00 or 11 are eliminated (depending on the initial density), we need 
to grow clusters of zeros or ones, in order to reach the final configuration of all 
all zeros or all ones. 

If, after multiple iterations of $F_X R_{184}$, every row included some 00 pairs (or some 11 pairs),
iterating  $R_{232}$ would produce the desired final configuration of all zeros (or all ones).
However, it is entirely possible that after iterations of $F_X R_{184}$ we will obtain some rows
which are made of alternating zeros and ones. $\ldots 010101\ldots$. Such rows after one iteration of rule 232
become $\ldots101010\ldots$, and after another iteration become again $\ldots 010101\ldots$, thus they
are periodic points of rule 232 (with period 2). This means that the desired final configuration of all zeros
or all ones may never be reached if the ``plain'' $R_{232}$ is iterated. We need a way to move $00$ and
$11$ pairs between rows, just like before. 

\begin{figure}
\begin{center}
 \begin{tikzpicture}[->,>=stealth',shorten >=1pt,auto,node distance=1.2cm,
 % thick,main node/.style={circle,fill=blue!20,draw,font=\sffamily\Large\bfseries}]
   thick,main node/.style={font=\bfseries}]

  \node[main node] (1)              {0};
  \node[main node] (2) [right of=1] {0};
  \node[main node] (3) [right of=2] {0};
  
  \node[main node] (4) [below of=1] {$\star$};
  \node[main node] (5) [right of=4] {1};
  \node[main node] (6) [right of=5] {$\star$};
  
  \node[main node] (7) [below of=4] {$\star$};
  \node[main node] (8) [right of=7] {$\star$};
  \node[main node] (9) [right of=8] {$\star$};

  \path[every node/.style={font=\sffamily}]
           (5)  edge [bend right] node[right] {$p=0.5$} (2);
\end{tikzpicture}
\quad \quad \quad
\begin{tikzpicture}[->,>=stealth',shorten >=1pt,auto,node distance=1.2cm,
 % thick,main node/.style={circle,fill=blue!20,draw,font=\sffamily\Large\bfseries}]
   thick,main node/.style={font=\bfseries}]

  \node[main node] (1)              {$\star$};
  \node[main node] (2) [right of=1] {0};
  \node[main node] (3) [right of=2] {$\star$};
  
  \node[main node] (4) [below of=1] {1};
  \node[main node] (5) [right of=4] {1};
  \node[main node] (6) [right of=5] {1};
  
  \node[main node] (7) [below of=4] {$\star$};
  \node[main node] (8) [right of=7] {$\star$};
  \node[main node] (9) [right of=8] {$\star$};

  \path[every node/.style={font=\sffamily}]
           (5)  edge [bend right] node[right] {$p=0.5$} (2);
\end{tikzpicture}
\end{center}
\caption{Particle (agent) representation of rule $G_X$.}\label{figgx}
\end{figure}
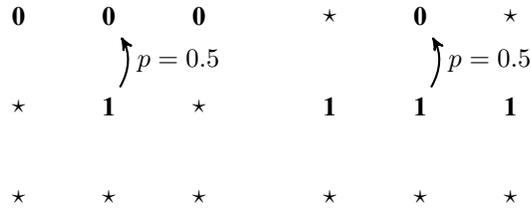

This will be done using a probabilistic rule $G_X$, slightly different that previously defined $F_X$. Using again
the agent paradigm, we will construct a rule in which an agent having 000 above itself will jump up
with probability 50\% or stay in the same place with probability 50\%. Similarly, an empty site
which has 111 below will become occupied with probability 50\% (the central agent from 111 below will
jump to it), and will remain unchanged with probability 50\%. This is illustrated in Figure~\ref{figgx}.
$G_X$ could be called a ``crowd avoidance'' rule, since agents tend to avoid being surrounded by other
agents, and tend to move into cells which have empty neighbours.
%%%%%%%%%%%%%%Gx

Using the notation introduced earlier,  we can define $G_X: \mathcal S \to \mathcal S$ as
\begin{equation}
[G_X(\mathbf{s})]_{i,j}=\left\{
\begin{array}{ll}
 1 & \mathrm{if\,\,}  \mathbf{u}_{i,j} =\left(
 {\setlength{\arraycolsep}{3pt}
  \begin{array}{lll}
  \star & \star & \star\\
  0 & 0 & 0\\
  \star & 1 & \star
  \end{array}}\right) \mathrm{and\,\,} X_{i,j-1}=1,\\
 0 & \mathrm{if\,\,}  \mathbf{u}_{i,j} =\left(
  {\setlength{\arraycolsep}{3pt}
  \begin{array}{lll}
  0 & 0 & 0\\
  \star & 1 & \star\\
  \star & \star & \star
  \end{array}}\right) \mathrm{and\,\,} X_{i,j}=1,\\
%%%%%%%%%%%
 1 & \mathrm{if\,\,}  \mathbf{u}_{i,j} =\left(
  {\setlength{\arraycolsep}{3pt}
  \begin{array}{lll}
  \star & \star & \star\\
  \star & 0 & \star\\
  1 & 1 & 1
  \end{array}}\right) \mathrm{and\,\,} X_{i,j-1}=1,\\
 0 & \mathrm{if\,\,}  \mathbf{u}_{i,j} =\left(
  {\setlength{\arraycolsep}{3pt}
  \begin{array}{lll}
  \star & 0 & \star\\
  1 & 1 & 1\\
  \star & \star & \star
  \end{array}}\right) \mathrm{and\,\,} X_{i,j}=1,\\	 
%%%%%%%%%%%%%	 
  s_{i,j}  & \mathrm{otherwise.}
\end{array}
\right.
\end{equation}
The above rule can be written in an algebraic form similarly as done
for rule $F_X$ in eq. (\ref{ruleFformula}). Then one can write
equation analogous to eq. (\ref{ruleFsum}) and show that $G_X$ 
is number-conserving, by shifting dummy indices in the same fashion
as we did for rule $F_X$. We  omit these details since they are 
rather straightforward but tedious. Moreover, Figure~\ref{figgx} makes it clear
that the number of agents must stay constant.

Rule $G_X R_{232}$ does exactly what is needed: it grows clusters of zeros or ones, but it also
eliminates rows of the type $\ldots 01010101 \ldots$. 
\section{Classification of densities}
The density classification can now be performed by
\begin{equation}
 (G_X R_{232})^{T_2} (F_X R_{184})^{T_1},
\end{equation}
where, obviously, we need to use different random field $X$ each
time step. More precisely, we define
\begin{multline}
(G_X R_{232})^{T_2}(F_X R_{184})^{T_1}=
  \left(F_{X^{(T_1+T_2)}} R_{232} \right)\ldots \left(F_{X^{(T_1+2)}} R_{232}\right)\\ 
\left(F_{X^{(T_1+1)}} R_{232}\right)   
 \left(F_{X^{(T_1)}} R_{184} \right)\ldots \left(F_{X^{(2)}} R_{184}\right) \left(F_{X^{(1)}} R_{184}\right)\\
 \end{multline}
where $X^{(1)}, X^{(2)}, \ldots, X^{(T_1+T_2)}$ are independent identically 
distributed random arrays, each consisting of random variables
 $X^{(t)}_{i,j}$ with probability distribution described by eq. (\ref{Xprobdist}).
To simplify our classification scheme, we will take $T_1=T_2$, even though this is
not the most efficient choice (typically, $T_2$ does not need to be as large as $T_1$).

Let $\mathbf{0}$ denote an $L \times L$ array of zeros, and $\mathbf{1}$
 an $L \times L$ array of ones. We propose the following conjecture.\\
\textbf{Conjecture.} %\begin{conjecture}
{ \it
Let $\mathbf{s}$ be an $L \times L$ array of binary numbers.
If the number of ones in $\mathbf{s}$ is greater than the number of zeros,
then, with probability 1,
\begin{equation}
 \lim_{T \to \infty} (G_X R_{232})^{T} (F_X R_{184})^{T}(\mathbf{s}) = \mathbf{1}.
\end{equation}
Similarly, if the number of ones in $\mathbf{s}$ is smaller than the number of zeros, then, with 
probability 1,
\begin{equation}
 \lim_{T \to \infty} (G_X R_{232})^{T} (F_X R_{184})^{T}(\mathbf{s}) = \mathbf{0}.
\end{equation}}
%\end{conjecture}
In order to provide numerical evidence supporting this conjecture, we will 
evaluate performance of the rule  $(G_X R_{232})^{T} (F_X R_{184})^{T}$ in classifying densities.
Of course, we won't be able to perform infinitely many
iterations. Nevertheless,  as we will see in the next section, the performance can be very 
good even if a finite $T$ is used.
\section{Performance}
{\em Performance} of a given rule $\Psi$ in performing the density
classification is typically defined as follows. Let $I$ denotes the number of
random initial configurations consisting of $N$ cells each, drawn from a symmetric
Bernoulli distribution. This means that each initial configuration is generated
by setting each of its cells independently to 0 or 1, with the same probability
$1/2$.  If the resulting configuration has exactly the same number of zeros and ones, we flip 
one randomly selected bit to break the symmetry.

 Suppose that we iterate the rule $\Psi$ on each of those initial
configurations. If a configuration with
initial density less than 0.5  converges to $\mathbf{0}$, we
consider it a successful classification, similarly as when a configuration with
initial density greater than 0.5 converges to $\mathbf{1}$. In
all other cases we consider the classification unsuccessful. The percentage of
successful classifications among all $I$ initial conditions will be 
 called \textit{performance} of the rule $\Psi$. 
\begin{table}
 \begin{center}
\begin{tabular}{c|c|c}
  & performance for & performance for \\ 
$T$  & $L=50$ & $L=100$\\ \hline
250 & 85.5\% & 39.9\% \\
500 & 96.1\% & 75.9\%\\
1000 & 99.4\% & 90.3\%\\
2000 & 99.7\% & 97.5\%\\	
4000 & 100\%  & 99.7\%\\
8000 & 100\%  & 99.9\% \\
16000 & 100\% & 100\%  
 \end{tabular}
 \end{center}
\caption{Performance of $(G_X R_{232})^{T} (F_X R_{184})^{T}$ for $I=100$ and for lattices 
$50 \times 50$ and $100 \times 100$ for different values of $T$.}
\end{table} 
In Table 1, we show the performance for $I=1000$ and for two array sizes,
$N=50 \times 50$ and $N=100 \times 100$. In order to show how the performance
depends on $T$, we used different values of $T$ ranging from 250 to 16000.
One can see that for $L=50$, 4000 iterations of each rule suffice to obtain
the perfect performance. For $L=100$, 16000 iterations are needed. This means that
one needs only slightly more than three iterations per bit in both of those cases.
This is understandable if one considers the ``diffusive'' nature of $F_X$ and $G_X$. 
In $F_XR_{184}$, for example, pairs $00$ perform a sort of random walk until they hit
a pair $11$. The required number of iterations of  $F_XR_{184}$, therefore, should be
strongly correlated with the average hitting time for random walk.
It is known that the average hitting time for a random walk on 2D periodic lattice
roughly scales as $L^2$ \cite{paper13}, and, since the number of bits in the lattice also scales as
 $L^2$,  we expect the the number of iterations per bit should remain roughly constant
as $L$ increases. 

We should also remark that although we took the number of iterations to be the same for both
rules $F_X R_{184}$ and $G_X R_{232}$, this is not really necessary. In fact, the largest 
number of iterations of $G_X R_{232}$ needed to converge to $\mathbf{0}$ or $\mathbf{1}$ 
observed in our numerical simulations  was 611 for $100 \times 100$ lattice and 265 for $50
\times 50$ lattice. Since clusters of zeros or ones in rule 232 grow linearly with time,
and they grow only in horizontal direction, one can expect that the number of required
 iterations of  $G_X R_{232}$ grows linearly with $L$.
\section{Conclusions and further work}
We presented construction of a probabilistic two-rule scheme which performs density classification
in two dimensions with increasing accuracy as the number of iterations increases. 
Although right now this scheme remains a conjecture, numerical evidence strongly
supports it. Moreover, since the dynamics of the classification process is rather well
understood, it is quite likely that a formal proof of this conjecture may be within reach.
It should also be possible to obtain some rigorous bounds on the expected number of iterations
needed for classification. 

A crucial feature of the proposed solution is that the agents at some point of time
change their behaviour from traffic coupled with lane changing to proliferation/death coupled
with crowd avoidance. Could one devise a single rule achieving similar performance?
In the one-dimensional case, this has been achieved, as the probabilistic
solution proposed in \cite{Fates2011} is actually a stochastic ``mixture'' of rules 184 and 232.
The author plans to pursue this idea in the near future.

One should also add that if one follows the spirit of the ``classical'' DCP, the agents should have
no access to any global information. Since in the proposed solution they change their behaviour
after $T$ iterations, one could argue that they have access to a global timer. At the same
time, the agents follow very simple rules, which makes it possible to describe their
dynamics as CA. This brings an interesting question: if one denied the agents the access to the
global timer, but instead equipped them with more sophisticated rules of behaviour (e.g., giving them
local memory, making them ``smarter'', etc.), could the DCP be solved? A~recent progress on solving DCP using CA with memory 
\cite{Alonso2013}  indicates that this could be a promising avenue to pursue. On might, therefore,
ask a broader and more general question: what are the least-complex agents which could solve the DCP? The author
hopes that this article stimulates further research in this direction.

%\balance
\section*{Acknowledgments}
The author acknowledges financial support from the Natural Sciences and
Engineering Research Council of Canada (NSERC) in the form of Discovery Grant.
This work was made possible by the facilities of the Shared
Hierarchical Academic Research Computing Network (SHARCNET:www.sharcnet.ca) and
Compute/Calcul Canada.

% serves to balance the column lengths on the last page of the document
% should be inserted the left column of the last page

\bibliography{JCA_0180_Fuks.bib} 

\begin{thebibliography}{10}

\bibitem{Alonso2013}
Ramón Alonso-Sanz and Larry Bull.
\newblock (2009).
\newblock A very effective density classifier two-dimensional cellular
  automaton with memory.
\newblock {\em Journal of Physics A: Mathematical and Theoretical},
  42(48):485101.

\bibitem{EJP2325}
Ana Bušić, Nazim Fatès, Jean Mairesse, and Irène Marcovici.
\newblock (2013).
\newblock Density classification on infinite lattices and trees.
\newblock {\em Electron. J. Probab.}, 18:no. 51, 1--22.

\bibitem{oliv2013}
Pedro~P.B de~Oliveira.
\newblock (2013).
\newblock Conceptual connections around density determination in cellular
  automata.
\newblock In Jarkko Kari, Martin Kutrib, and Andreas Malcher, editors, {\em
  Cellular Automata and Discrete Complex Systems}, volume 8155, pages 1--14.
  Springer Berlin Heidelberg.

\bibitem{Fates2011}
N.~Fat{\`e}s.
\newblock (2011).
\newblock {Stochastic Cellular Automata Solve the Density Classification
  Problem with an Arbitrary Precision}.
\newblock In Thomas Schwentick and Christoph D{\"u}rr, editors, {\em 28th
  International Symposium on Theoretical Aspects of Computer Science (STACS
  2011)}, volume~9, pages 284--295, Dagstuhl, Germany. Schloss
  Dagstuhl--Leibniz-Zentrum fuer Informatik.

\bibitem{Fates2d-2011}
N.~Fatès.
\newblock (2012).
\newblock A note on the density classification problem in two dimensions.
\newblock In E.~Formenti, editor, {\em Proc. of the 18th International Workshop
  on Cellular Automata and Discrete Complex Systems: Exploratory Papers
  Proceedings}, pages 11--18. Rapport de Recherche I3S - ISRN:
  I3S/RR-2012-04-FR.

\bibitem{Fuks97}
H.~Fuk{\'s}.
\newblock (1997).
\newblock Solution of the density classification problem with two cellular
  automata rules.
\newblock {\em Phys. Rev. E}, 55:2081R--2084R.

\bibitem{paper10}
H.~Fuk{\'s}.
\newblock (2000).
\newblock A class of cellular automata equivalent to deterministic particle
  systems.
\newblock In A.~T.~Lawniczak S.~Feng and R.~S. Varadhan, editors, {\em
  Hydrodynamic Limits and Related Topics}, Fields Institute Communications
  Series, Providence, RI. AMS.

\bibitem{paper20}
H.~Fuk{\'s}.
\newblock (2002).
\newblock Non-deterministic density classification with diffusive probabilistic
  cellular automata.
\newblock {\em Phys. Rev. E}, 66:066106.

\bibitem{paper13}
H.~Fuk{\'s}, A.~T. Lawniczak, and S.~Volkov.
\newblock (2001).
\newblock Packet delay in models of data networks.
\newblock {\em ACM Transactions on Modelling and Simulations}, 11:233--250.

\bibitem{Gacs78}
P.~Gacs, G.~L. Kurdymov, and L.~A. Levin.
\newblock (1987).
\newblock One-dimensional uniform array that wash out finite islands.
\newblock {\em Probl. Peredachi Inform.}, 14:92--98.

\bibitem{land95}
M.~Land and R.~K. Belew.
\newblock (1995).
\newblock No perfect two-state cellular automata for density classification
  exists.
\newblock {\em Phys. Rev. Lett.}, 74(25):5148--5150.

\bibitem{oliv2005}
ClaudioL.M. Martins and PedroP.B. Oliveira.
\newblock (2005).
\newblock Evolving sequential combinations of elementary cellular automata
  rules.
\newblock In MathieuS. Capcarrère, AlexA. Freitas, PeterJ. Bentley, ColinG.
  Johnson, and Jon Timmis, editors, {\em Advances in Artificial Life}, volume
  3630, pages 461--470. Springer Berlin Heidelberg.

\bibitem{JimenezMorales2001571}
F.~Jiménez Morales, J.P. Crutchfield, and M.~Mitchell.
\newblock (2001).
\newblock Evolving two-dimensional cellular automata to perform density
  classification: A report on work in progress.
\newblock {\em Parallel Computing}, 27(5):571--585.

\bibitem{Wolf2008}
D.~Wolz and P.~P.~B. de~Oliveira.
\newblock (2008).
\newblock Very effective evolutionary techniques for searching cellular
  automata rule spaces.
\newblock {\em Journal of Cellular Automata}, 3(4):289--312.

\end{thebibliography}
\end{document}